\documentclass{optica-article}

\journal{opticajournal} 

\articletype{Research Article}
\usepackage{amsbsy}
\usepackage{bibunits}
\usepackage{babel}
\usepackage{amsmath, nccmath}
\usepackage{bm}
\usepackage{algorithm2e}
\usepackage{lineno}
\usepackage{physics}
\usepackage{empheq}
\usepackage[]{graphicx}
\usepackage{enumitem}
\usepackage{booktabs}
\newcommand*\widefbox[1]{\fbox{\hspace{0em}#1\hspace{0em}}}

\begin{document}

\title{What are the quantum commutation relations for the total angular momentum of light?}

\author{Pronoy Das,\authormark{1} Li-Ping Yang,\authormark{2} and Zubin Jacob\authormark{1,*}}

\address{\authormark{1}Elmore Family School of Electrical and Computer Engineering, Birck Nanotechnology Center, Purdue University, West Lafayette, IN 47907, United States of America\\
\authormark{2}Center for Quantum Sciences and School of Physics, Northeast Normal University, Changchun 130024, China\\
}

\email{\authormark{*}zjacob@purdue.edu} 

\begin{abstract*} 
The total angular momentum of light has received  attention for its application in
a variety of phenomena such as optical communication, optical forces and sensing. However, the quantum behavior including the commutation relations have been relatively less explored. Here, we derive the correct commutation relation for the total angular momentum of light using both relativistic and non-relativistic approaches. An important outcome of our work is the proof that the widely-assumed quantum commutation relation for the total observable angular momentum of light is fundamentally incorrect. Our work will motivate experiments and leads to new insight on the quantum behavior of the angular momentum of light.

\end{abstract*}

\section{Introduction}

A photon possesses two distinct types of angular momenta (AM): the spin angular momentum (SAM) and the orbital angular momentum (OAM). The rotation of the polarization vector defines the SAM, and the twist of the phase-front defines the OAM of the photon. Poynting \cite{Poynting1909} in 1909 pioneered the concept of the AM in a polarized light beam, later corroborated by Raman \cite{Raman1931} (1931) and Darwin \cite{Darwin1932} (1932). The first experimental evidence of the AM of light is accredited to the mechanical detection of the AM in the seminal experiment by Beth \cite{Beth1936} in 1936. A few decades later, Allen \textit{et. al.} \cite{Allen1992} in 1992 showed that light with a helical phase-front carry OAM. This discovery propelled the advancements exploiting the OAM of light for various applications \cite{Dunlop2017}, such as high-bandwidth information encoding for enhanced optical communication \cite{Zhou2021, Zeilinger16, Willer18, Willner15} and oriental trapping and manipulation of particles using optical tweezers \cite{Friese1996} which is extensively used in biophysics \cite{Carlos2021, Ashkin1987}.


However, the majority of the advancements are limited to the semi-classical interpretation of the total angular momentum of the photon and the ensuing commutation relations, and the realm of quantum for the AM remains relatively uncharted \cite{Yang2021, Das2024arxiv}. This paper presents a thorough examination of the widely-assumed commutation relations for the angular momentum of light, along with the rigorous derivation of the correct quantum commutation relations. Our derivations encompass both the non-relativistic \cite{Mandel1982,Enk1994} and relativistic domains \cite{Yang2022,Yang2021} to demonstrate the consistency of the derived commutation relations. We note that the incorrect commutation relations occur in a vast amount of literature.
\begin{empheq}[box=\widefbox]{align*}
    &[\hat{J}_i,\hat{J}_j] =i\hbar \varepsilon_{ijk}\hat{L}_k &\text{(Correct)}\\
    &[\hat{L}_i,\hat{L}_j] =i\hbar \varepsilon_{ijk}\hat{L}_k &\text{(Correct)}\\
    &[\hat{L}_i,\hat{S}_j] =0 &\text{(Correct)}
\end{empheq}
\begin{align*}
    &[\hat{J}_i,\hat{J}_j]=i\hbar \varepsilon_{ijk}\hat{J}_k &\text{(Incorrect)}\\ 
    &[\hat{L}_i,\hat{L}_j]=i\hbar \varepsilon_{ijk}(\hat{L}_k + \hat{S}_k)&\text{(Incorrect)}\\
    &[\hat{L}_i,\hat{S}_j]=i\hbar \varepsilon_{ijk}\hat{S}_k &\text{(Incorrect)}
\end{align*}

\section{Non-relativistic approach for AM commutation relations}

The total angular momentum of a classical electromagnetic field is given by \cite{Mandel1982, Tannoudji1997}:
\begin{equation}
    \hat{\bm{J}}=\varepsilon_0 \int d^3r\;\bm{r} \times (\hat{\bm{E}}\times \hat{\bm{B}}),
    \label{eq1}
\end{equation}
where, $\hat{\bm{E}}$ and $\hat{\bm{B}}$ are the electric and magnetic field vectors respectively, and $\bm{r}$ is the position vector. In a fixed coordinate system, the field vectors exhibit gauge independence. We employ Helmholtz decomposition on the vector fields to separate the rotational (transverse) and non-rotational (longitudinal) components, and discard the non-rotational component under Coulomb gauge fixing conditions \cite{Tannoudji1997} (Appendix A). The gauge invariant expression of the AM is given by:
\begin{equation} 
    \begin{split}
        \hat{\bm{J}}=\left(\underbrace{\varepsilon_0\int d^3r [\hat{\bm{E}}_\perp (\bm{r} \times \nabla)\hat{\bm{A}}_\perp]}_\text{orbital angular momentum} \right) + \left(\underbrace{\varepsilon_0\int d^3r[\hat{\bm{E}}_\perp \times \hat{\bm{A}}_\perp]}_\text{spin angular momentum} \right).
    \end{split}
    \label{eq2}
\end{equation}

\noindent The first term is defined as the orbital angular momentum (OAM, $\bm{\hat{L}}$) and the second term is the spin angular momentum (SAM, $\bm{\hat{S}}$) of the field.

The plane-wave expansion of the field operators in free space can be given by (considering the circularly polarized plane waves as our field modes) \cite{Enk1994}:
\begin{align}
    \bm{\hat{A}}_\perp (\bm{r},t)&= \int d^3k \sum_{\lambda=\pm 1}\sqrt{\frac{\hbar}{2\varepsilon_0 \omega}} \left(\hat{a}_{\bm{k},\lambda} \bm{\epsilon}(\bm{k},\lambda) e^{i(\bm{k.r}-\omega t)} +h.c. \right), \\
    \bm{\hat{E}}_\perp (\bm{r},t)&= i\int d^3k \sum_{\lambda=\pm 1} \sqrt{\frac{\hbar \omega}{2\varepsilon_0}} \left(\hat{a}_{\bm{k},\lambda} \bm{\epsilon}(\bm{k},\lambda) e^{i(\bm{k.r}-\omega t)} - h.c. \right),
    \label{eq3}
\end{align}
where $\varepsilon_0$ is the permittivity in vacuum and $\omega = c|\bm{k}|$ is the mode frequency. $\bm{\epsilon}(\bm{k},\lambda)$ are the transverse unit circular-polarization vectors. We can write these unit vectors in terms of the linear polarization vectors as: 
\begin{equation}
    \bm{\epsilon}(\bm{k},\lambda)=\frac{1}{\sqrt{2}}\Big( -\lambda \bm{\epsilon}(\bm{k},1) + i \bm{\epsilon}(\bm{k},2)  \Big), \quad \lambda=\pm 1.
\end{equation}
These polarization unit vectors follow the transversality relation $\bm{k} \cdot \bm{\epsilon}(\bm{k},\lambda)=0$, and the orthonormality relation $\epsilon^*(\bm{k},\lambda)\cdot\bm{\epsilon}(\bm{k}',\lambda')=\delta^3(\bm{k-k'})\delta_{\lambda,\lambda'}$. The unit vector along the longitudinal direction is along $\bm{k}$, given by $\bm{\epsilon}(\bm{k},3)=\hat{\bm{k}}=\bm{k}/k$. 

The operators $\hat{a}^\dagger_{\bm{k},\lambda}$ and $\hat{a}_{\bm{k},\lambda}$ are the creator and annihilator operators of the bosonic modes of the photon, and follow the commutation relations:
\begin{equation}
    [\hat{a}_{\bm{k},\lambda},a^\dagger_{\bm{k}^\prime,\lambda^\prime}]=\delta^3(\bm{k-k}^\prime)\delta_{\lambda,\lambda^\prime},\quad [\hat{a}_{\bm{k},\lambda},a_{\bm{k}^\prime,\lambda^\prime}]=[\hat{a}^\dagger_{\bm{k},\lambda},a^\dagger_{\bm{k}^\prime,\lambda^\prime}]=0.
\end{equation}
Using these relations, we perform the canonical quantization of the AM under Coulomb gauge conditions and derive the following commutation relations (Appendix A.1):
\begin{align}
    &[\hat{L}_i,\hat{L}_j]=i\hbar \varepsilon_{ijk}\hat{L}_k,\\
    &[\hat{L}_i,\hat{S}_j]=0,\\
    &[\hat{J}_i,\hat{J}_j]=i\hbar \varepsilon_{ijk} \hat{L}_k,
\end{align}
and $[\hat{S}_i,\hat{S}_j]=0$ (this specific SAM commutation relation is consistent with \cite{Enk1994}). Our calculations show that the gauge-invariant AM operator does not conform to the commutation relations associated with the generator of rotations in the SO(3) group (since $ [\hat{J}_i,\hat{J}_j]\neq i\hbar \varepsilon_{ijk}\hat{J}_k $). Also, the commutation relations of the SAM and OAM reveal that they are independently observable quantities.

To further substantiate this result, we consider a general field vector $\bm{\hat{V}}(\bm{r},t)$ in the SO(3) group, and investigate whether the vector $\bm{\hat{J}}$ is a generator of rotations for $\bm{\hat{V}}$. We decompose this field vector into the transverse and longitudinal components as per Helmholtz decomposition:
\begin{equation}
    \bm{\hat{V}}^\perp(\bm{r},t)= \int d^3k \mathcal{N}(\bm{k}) [\hat{a}_{\bm{k},-1}\bm{\epsilon}(\bm{k},-1)e^{i(\bm{k.r}-\omega t)} + \hat{a}_{\bm{k},+1}\bm{\epsilon}(\bm{k},+1)e^{i(\bm{k.r}-\omega t)} + h.c.],
\end{equation}
\begin{equation}
    \bm{\hat{V}}^\parallel(\bm{r},t)= \int d^3k \mathcal{N}(\bm{k}) [\hat{a}_{\bm{k},3}\bm{\epsilon}(\bm{k},3)e^{i(\bm{k.r}-\omega t)} + h.c.],
\end{equation}
where $\mathcal{N}(\bm{k})$ is the normalization constant. If we consider an operator $\hat{\mathcal{R}}$ that generates rotations preserving the SO(3) group structure, it will follow the condition:
\begin{equation}
    \hat{V}'_i=\hat{V}_i+\frac{i}{\hbar}[\hat{V}_i,\hat{\mathcal{R}}_j]\delta \theta_j,
    \label{eq12}
\end{equation}
where $[\hat{V}_i,\hat{\mathcal{R}}_j]$ has the form: $[\hat{V}_i,\hat{\mathcal{R}}_j]=i\hbar\varepsilon_{ijk}\hat{V}_k+$ (other norm-preserving rotational terms), and $\delta \theta_j$ is an infinitesimally small rotation along $j$. Evaluating the commutator between the general field $V(\bm{r},t)$ and the quantum AM operators we have [Appendix A.2]:
\begin{align}
    &[\hat{J}_i,\hat{V}_j^\perp]=i\hbar\left( \varepsilon_{ijk}\hat{V}_k^\perp - \varepsilon_{ink}r'_n \nabla_k \hat{V}_j^\perp \right),\\
    &[\hat{J}_i,\hat{V}^\parallel]=[\hat{L}_i,\hat{V}^\parallel]=[\hat{S}_i,\hat{V}^\parallel]=0.
    \label{eq14}
\end{align}
Here, the term $i\hbar\varepsilon_{ijk}\hat{V}_k^\perp$ defines the intrinsic `twist' of the field vector by the SAM operator, and the $i\hbar\varepsilon_{ink}r'_n \nabla_k \hat{V}_j^\perp$ term represents the rotation of the field vector by the OAM operator about $\bm{r}$. From Equations \ref{eq12}, 13 and \ref{eq14} we prove that for a general vector field $\hat{V}(\bm{r},t)$ in SO(3) under Helmholtz decomposition, only the transverse component of the field $\hat{V}^\perp (\bm{r},t)$ rotates as a vector under $\bm{\hat{J}}$. Thus, a general field vector in SO(3) does not rotate as a vector under $\bm{\hat{J}}$.

We can reduce Equation 13 to the form \cite{Mandel1982}:
\begin{equation}
    \hat{\bm{V}}^{\perp \prime}(\bm{r},t)=R_{(2)} \hat{\bm{V}}^\perp(R_{(2)}^{-1}\bm{r},t),
\end{equation}
where $R_{(2)}$ is the rotation matrix that rotates a vector by $\delta \bm{\theta}$ normal to $\bm{k}$. This statement proves that the AM operator $\hat{\bm{J}}$ is a generator of rotations for observable fields  $\hat{\bm{V}}^\perp(\bm{r},t)$ (such as $\hat{\bm{E}}$ and $\hat{\bm{B}}$ fields). These rotations do not lie in the SO(3) group, but they exist in a subgroup of SO(3), where the rotations are confined to the plane normal to $\bm{k}$. Hence, they do not follow the property for the generators of the SO(3) rotation group ($[\hat{J}_i,\hat{J}_j] \neq i\hbar \varepsilon_{ijk}\hat{J}_k$).

We note that, at first glance, the OAM operator appears to be the generator of rotations in SO(3), since it follows the commutation relation $[\hat{L}_i,\hat{L}_j]=i\hbar \varepsilon_{ijk}\hat{L}_k$. Yet, we have shown in Appendix A that $[\hat{J}_i,\hat{V}^\parallel_j]=[\hat{S}_i,\hat{V}^\parallel_j]=0 \Rightarrow [\hat{L}_i,\hat{V}^\parallel_j]=0$, meaning, if we perform an infinitesimal rotation $\delta \bm{\theta}$ of a general vector field $\bm{\hat{V}}(\bm{r},t)\rightarrow \bm{\hat{V}}'(\bm{r},t)$, we get:
\begin{align*}
    \hat{V}'_i=&\hat{V}_i+\frac{i}{\hbar}[\hat{V}_i,\hat{L}_j]\delta \theta_j \\
    =& \hat{V}_i+\frac{i}{\hbar}\left([\hat{V}^\perp_i,\hat{L}_j] + [\hat{V}^\parallel_i,\hat{L}_j]\right)\delta \theta_j \\
    =& \hat{V}_i + \frac{i}{\hbar}[\hat{V}^\perp_i,\hat{L}_j]\delta \theta_j= \hat{V}_i + \frac{i}{\hbar}[\hat{V}^\perp_i,\hat{J}_j]\delta \theta_j,
\end{align*}
where, only the transverse component undergoes a rotation, proving $\bm{\hat{L}}$ does not preserve the SO(3) group structure. The commutation relations are necessary, but not a sufficient condition for the generator of rotations in SO(3).







\section{Relativistic approach for AM commutation relations}
In this section, we exploit the symmetries of the SO(3) group and obtain the generators of the group as the angular momentum operators using Noether's theorem. We note that these AM operators are gauge-dependent under U(1), hence we decompose the field vectors to remove the gauge-dependent components to arrive at the commutation relations for the observable AM. 

In Appendix B, we derive the AM operators using Noether's theorem, given by \cite{Greiner1993, Yang2022}:
\begin{equation} 
    \begin{split}
        \hat{\bm{\Tilde{J}}}=
        \left(\underbrace{\frac{1}{c}\int d^3r [-\hat{\pi}^\mu(\bm{r} \times \nabla ) \hat{A}_\mu]}_{\text{OAM},\hat{\bm{\Tilde{L}}}} \right) + \left(\underbrace{\frac{1}{c}\int d^3r \bm{\hat{\pi}} \times \bm{\hat{A}}}_{\text{SAM},\hat{\bm{\Tilde{S}}}} \right),
    \end{split}
    \label{eq15}
\end{equation}
where  $\bm{\hat{\pi}}=-(1/\mu_0)\partial_0 A^\mu$ is the conjugate momentum of the gauge field $\bm{\hat{A}}$. Since these operators are not gauge-invariant under U(1), we distinguish them from the rest using a \textit{tilde}. The plane-wave expansion of the field operators in terms of the bosonic mode operators are:
\begin{align}
    \hat{A}^{\mu} & =\int d^{3}k\sum_{\lambda=0}^{3}\sqrt{\frac{\hbar}{2\varepsilon_{0}\omega_k(2\pi)^{3}}}[\hat{a}_{\bm{k},\lambda}\epsilon^{\mu}(\boldsymbol{k},\lambda)e^{ik^\nu r_\nu}+h.c.],\\
    \hat{\pi}^{\mu} & =i\int d^{3}k\sum_{\lambda=0}^{3}\sqrt{\frac{\hbar\omega_k}{2\mu_{0}(2\pi)^{3}}}[\hat{a}_{\bm{k},\lambda}\epsilon^{\mu}(\boldsymbol{k},\lambda)e^{ik^\nu r_\nu}-h.c.],
\end{align}
where the bosonic mode ladder operators follow the bosonic commutation relations. $\epsilon^\mu(\bm{k},\lambda)$ is the polarization unit four-vector, and follow the orthonormality condition $\epsilon_\mu(\bm{k},\lambda)\epsilon^\mu(\bm{k},\lambda')=g_{\lambda,\lambda'}$. The longitudinal unit vector $\lambda=3$ is parallel to $\bm{k}$ and thus given by $\epsilon(\bm{k},3)=(0,\bm{k}/|k|)$. The time-like polarization vector $\epsilon(\bm{k},0)=(1,0,0,0)$ represent the scalar photon.

We postulate the following Equal Time Commutation Relations (ETCRs) \cite{Yang2022} for the quantization of the AM operator:
\begin{align}
    [\hat{A}^\mu (\bm{r}),\hat{\pi}^\nu (\bm{r}',t)]=i\hbar c g^{\mu \nu} \delta^3 (\bm{r-r'}),\\
    [\hat{A}^\mu (\bm{r}), \hat{A}^\nu (\bm{r}')]=[\hat{\pi}^\mu (\bm{r}), \hat{\pi}^\nu (\bm{r}')]=0.
\end{align}
We use the ETCR relations above to derive the following commutation relations of the AM (Appendix B.2) \cite{Yang2022}:
\begin{equation}
    [\hat{\Tilde{J}}_i,\hat{\Tilde{J}}_j\;]=i\hbar \varepsilon_{ijk} \hat{\Tilde{J}}_k.
    \label{eq21}
\end{equation}
This exhibits a striking resemblance to the well-known commutation relations of AM for a Dirac field. Moreover, one can prove that this AM operator is a generator of rotations of the SO(3) group using a similar approach outlined in Section 2.

However, we acknowledge that $\hat{\Tilde{J}}$ accounts for all four polarization degrees of freedom. Since photons are massless particles, they always carry a non-zero spin angular momentum. Thus, a spin-less particle such as the scalar photon does not exist, i.e. $\lambda=0$. We note that Equation \ref{eq21} still holds true with the remaining three polarization degrees of freedom (Appendix C).

For chargeless fields such as the photon, from the Maxwell equations we get $\nabla\cdot \bm{\hat{E}} =0$. This nullifies the existence of the longitudinal vectors along $\epsilon^\mu(\bm{k},3)=\bm{k}/k$ for the fields under Helmholtz decomposition \cite{Tannoudji1997}. Excluding the longitudinal polarization degree of freedom from the vector fields, we obtain the gauge-invariant (observable) fields that contains only two polarization degrees of freedom (i.e. two space-like transverse unit polarization vectors, $\lambda=1,2$).

We decompose the AM of the photon to extract the gauge-invariant part out [Appendix B]. We prove that using these conditions, the commutation relations for the observable AM operators are:
\begin{align}
    &[\hat{S}_i, \hat{S}_j]=0, \quad \leftarrow\text{consistent with \cite{Enk1994}}\\
    &[\hat{L}_i,\hat{L}_j]=i\hbar \epsilon_\textit{ijk}\hat{L}_k,\\
    &[\hat{L}_i, \hat{S}_j]=0,\\
    \therefore &[\hat{J}_i,\hat{J}_j]=i\hbar \varepsilon_{ijk} \hat{L}_k.
\end{align}
\textbf{Thus, we obtain the commutation relations for the relativistic total AM operators that are consistent with the non-relativistic counterparts.}

\section{Quantum AM operators as the generator of rotations}

We start with the gauge-dependent AM operator derived from Noether's theorem in the previous section, omitting the case of the scalar photon ($\lambda=0$) since photons are massless. Considering an arbitrary gauge field in SO(3):
\begin{equation}
    {\hat{V}}^\mu(\bm{r},t)= \frac{1}{\sqrt{(2\pi)^3}}\int d^3k \mathcal{N}(\bm{k}) \sum_{\lambda=1}^3[\hat{a}_{\bm{k},\lambda}\epsilon^\mu(\bm{k},\lambda)e^{ik^\nu r_\nu} + h.c.],
\end{equation}
from Appendix C we get:
\begin{align}
    &[\hat{\tilde{L}}_i,\hat{V}_j(\bm{r},t)]=-i\hbar \epsilon_{imn}r_n \nabla_m\hat{V}_j,\\
    &[\hat{\tilde{S}}_i,\hat{V}_j(\bm{r},t)]=i\hbar \epsilon_{ijk} \hat{V}_k,\\
    \therefore &[\hat{\tilde{J}}_i,\hat{V}_j(\bm{r},t)]=i\hbar \left( \epsilon_{ijk} \hat{V}_k-\epsilon_{imn}r_n \nabla_m\hat{V}_j \right).
\end{align}
Here, the gauge-dependent OAM operator rotates the vector field $\hat{V}(\bm{r},t)$ about $\bm{r}$ and the gauge-dependent SAM operator is responsible for the intrinsic rotation of the vector field. For an infinitesimally small angle $\delta \bm{\theta}$, we show that the gauge field rotates as:
\begin{align}
    \hat{V}'(\bm{r},t)=\left(1+\delta\bm{\theta}\times\right)\hat{V}(\bm{r}-\delta{\bm{\theta}}\times\bm{r},t).
    \label{eq29}
\end{align}
Equation \ref{eq29} proves that we can obtain the rotated vector $\hat{V}'(\bm{r},t)$ by rotating the position vector $\bm{r}$ by $-\delta \bm{\theta}$, then rotate the entire field vector $\hat{V}$ by $+\delta \bm{\theta}$. In other words, we can symbolically write $\hat{V}'(\bm{r},t)$ as:
\begin{equation}
    \hat{V}'(\bm{r},t)= R_{(3)} \hat{V}(R_{(3)}^{-1}\bm{r},t),
    \label{eq31}
\end{equation}
where $R_{(3)}$ is the 3D rotation matrix that rotates a vector by $\delta \bm{\theta}$. Equation \ref{eq31} explicitly proves that the gauge-dependent total AM operator $\hat{\tilde{\bm{J}}}$ is a generator of rotations for an arbitrary gauge field in SO(3) \cite{Mehta1968}. This further justifies the commutation relation property for $\hat{\tilde{\bm{J}}}$, i.e. $[\hat{\Tilde{J}}_i,\hat{\Tilde{J}}_j\;]=i\hbar \varepsilon_{ijk} \hat{\Tilde{J}}_k$. Thus, $\hat{\tilde{\bm{J}}}$ is the true AM operator for the gauge field.

Following the Maxwell equations, we can decompose the pure gauge-dependent (longitudinal) degree of freedom out of the quantum AM operators, which confines the fields and the field rotations in a subgroup of SO(3) (Section 2). Yet, Equation \ref{eq15} proves that the gauge-independent total AM operator $\hat{\bm{J}}$ is a generator of rotations for the gauge-independent or observable fields such as the electric $\hat{\bm{E}}$ and the magnetic $\hat{\bm{B}}$ field vectors. Hence, the quantum operator $\hat{\bm{J}}$ is the true AM operator for the observable fields.
\begin{table}[!htbp]
    \centering
    \begin{tabular}{|l|l|l|}
    \multicolumn{1}{c}{\textit{Gauge fields} $\hat{V}(\bm{r},t)$ (this work)} &\multicolumn{1}{c}{}& \multicolumn{1}{c}{\textit{Observable fields} $\hat{\bm{V}}^\perp(\bm{r},t)$\cite{Mandel1982}}\tabularnewline
    \cline{1-1} \cline{3-3} 
    $[\hat{\tilde{L}}_{i},\hat{V}_{j}]=-i\hbar\epsilon_{imn}r_{m}\frac{\partial}{\partial r_{n}}\hat{V_{j}}$ && $[\hat{L}_{i},\hat{V}_{j}]=-i\hbar\epsilon_{imn}r_{m}\frac{\partial}{\partial r_{n}}\hat{V_{n}}^{\perp}$\\
    $[\hat{\tilde{S}}_{i},\hat{V}_{j}]=i\hbar\epsilon_{ijk}\hat{V_{r}}$ && $[\hat{S}_{i},\hat{V}_{j}]=i\hbar\epsilon_{ijk}\hat{V_{r}}^{\perp}$\\
    \cline{1-1} \cline{3-3}
    $[\hat{\tilde{J}}_{i},\hat{V}_{j}]=i\hbar\left(\epsilon_{ijk}\hat{V_{r}}-\epsilon_{imn}r_{m}\frac{\partial}{\partial r_{n}}\hat{V_{j}}\right)$ && $[\hat{J}_{i},\hat{V}_{j}]=i\hbar\left(\epsilon_{ijk}\hat{V_{r}}^{\perp}-\epsilon_{imn}r_{m}\frac{\partial}{\partial r_{n}}\hat{V_{j}}^{\perp}\right)$\\
     
    $\hat{V}'(\bm{r},t)= R_{(3)} \hat{V}(R_{(3)}^{-1}\bm{r},t)$ && $\hat{\bm{V}}'(\bm{r},t)= R_{(2)} \hat{V}(R_{(2)}^{-1}\bm{r},t)$\\
    \cline{1-1} \cline{3-3}
    \end{tabular}
    \caption{\textbf{Gauge and Observable field vectors under rotation:} Perfect symmetry of the rotational properties between the gauge field operators and the observable field operators. We show that the total quantum AM operators are the generator of rotations for their respective fields and their respective symmetry groups.}
    \label{comp_table}
\end{table}

\section{Conclusion}
In summary, our comprehensive analysis shows the inadequacy of the widely-assumed quantum commutation relations. We have established the correct quantum commutation relations in both non-relativistic and relativistic domains.

By unveiling the correct commutation relations, our work advances the fundamental understanding of the photons and the AM behavior in the single-photon regime. We unlock new paths for studying the interplay between SAM and OAM in quantum optics, condensed matter physics, and quantum information processing, where accurate characterization and manipulation of angular momentum states are crucial.

\section{Acknowledgements}
This work is supported by the funding from Army Research Office (W911NF-21-1-0287).
\vspace{1cm}

\section*{\fontsize{12}{12}\selectfont Appendix}
\medskip
\appendix

\section{Non-relativistic approach}

From the classical definition of AM:
The most familiar expression for classical electromagnetic field is:
\begin{equation}
    \hat{\bm{J}}=\varepsilon_0 \int d^3r  \bm{r} \times (\hat{\bm{E}}\times \hat{\bm{B}}).
\end{equation}
In Coulomb gauge, $\boldsymbol{\nabla} \cdot \bm{\hat{A}}=0$. If we transform this gauge as $\bm{\hat{A}}^\prime = \bm{\hat{A}} -\nabla ^2 f$, under Coulomb gauge fixing we have:
\begin{equation}
    \nabla ^2 f=\boldsymbol{\nabla} \cdot \hat{\bm{A}}.
\end{equation}
This has the form of the Poisson's equation, implying that there exists a unique solution for this equation of the form: 
\begin{equation}
    f(\bm{r},t)=\int d^3r^\prime G(\bm{r}-\bm{r}^\prime)\boldsymbol{\nabla}^\prime\cdot \hat{\bm{A}}(\bm{r}^\prime,t),
\end{equation}
were, $G(\bm{r}-\bm{r}^\prime)=-(1/4\pi)(1/|\bm{r}-\bm{r}^\prime|)$ is the Green's function for this Poisson's equation.

Under Helmholtz equation, the vector potential can be split into parallel and perpendicular components, with the properties:
\begin{equation}
    \boldsymbol{\nabla} \times \hat{\bm{A}}_\parallel = 0, \qquad \boldsymbol{\nabla} \cdot \hat{\bm{A}}_\perp=0.
\end{equation}    
\textbf{Note 1} : The equation above also tells us that unlike the term $\bm{A_\perp}$, $\bm{A_\parallel}$ is dependent on the choice of the gauge.

\noindent\textbf{Note 2:} Since $\boldsymbol{\nabla} \cdot \hat{\bm{A}}=0=\boldsymbol{\nabla} \cdot \hat{\bm{A}}_\parallel$, $\hat{\bm{A}}_\parallel$ has vanishing curl (i.e. no rotation) and divergence (i.e. no compression/expansion). Hence, it is a constant vector field in a Euclidean space. Since the field vanishes at infinity, $\hat{\bm{A}}_\parallel$ is equal to zero at all points.

Photons do not exhibit a charge distribution, thus $\boldsymbol{\nabla} \cdot \hat{\bm{E}}=0\Rightarrow\hat{\bm{J}}_\parallel=0$ \cite{Tannoudji1997}. For the transverse component of the AM $\hat{\bm{J}}_\perp \equiv \hat{\bm{J}}$, we have:

\begin{equation}
    \hat{\bm{J}}=\varepsilon_0 \int d^3r [\hat{\bm{E}}_\perp (\bm{r} \times \nabla)\hat{\bm{A}}_\perp + \hat{\bm{E}}_\perp \times \hat{\bm{A}}_\perp]
\end{equation}
Here,
\begin{align}
    \hat{\bm{L}}&=\varepsilon_0 \int d^3r [\hat{\bm{E}}_\perp (\bm{r} \times \nabla)\hat{\bm{A}}_\perp]\\
    \hat{\bm{S}}&=\varepsilon_0 \int d^3r [\hat{\bm{E}}_\perp \times \hat{\bm{A}}_\perp]
\end{align}

We can substitute the plane-wave expansions of the field vectors \ref{eq3} in the equations above and obtain the quantized AM operators.

\subsection{Commutation relations for the angular momentum operators}
Using the commutation relations of the bosonic mode ladder operators, we obtain:
\begin{align*}
    &[\bm{\hat{A}}_\perp (\bm{r},t),\hat{E}_\perp (\bm{r}',t)]\\
    &= \frac{i\hbar}{2\varepsilon_0}\int d^3k d^3k'\left[ \sum_\lambda \frac{1}{\sqrt{\omega_k}}\left(\hat{a}_{\bm{k},\lambda} \bm{\epsilon}(\bm{k},\lambda) e^{i(\bm{k.r}-\omega t)} +h.c. \right), \sum_{\lambda'} \sqrt{\omega_{k'}} \left(a_{\bm{k}',\lambda'} \epsilon( \bm{k}',\lambda') e^{i(\bm{k}'.\bm{r}-\omega' t')} - h.c. \right) \right] \\
    &= -\frac{i\hbar}\varepsilon_0 \delta^\perp_{ij}(\bm{r}-\bm{r}'),
\end{align*}
where, $\delta^\perp_{ij}(\bm{r}-\bm{r}')$ is the projection of the delta function on the transverse plane \cite{Enk2007}. Similarly, it is trivial to prove:
\begin{equation*}
    [\hat{A}_{\perp,i} (\bm{r}),\hat{A}_{\perp,j} (\bm{r}')]=[\hat{E}_{\perp,i} (\bm{r}),\hat{E}_{\perp,j} (\bm{r}')]=0.
\end{equation*}
Using these commutation relations, we can explicitly prove the following commutation relations:
\begin{align*}
    \left[\hat{L}_i, \hat{L}_{j}\right] =& \varepsilon_0^2 \int d^3r\int d^3r^{\prime}\left[\hat{E}_{\perp,m} (\bm{r},t)(\bm{r} \times \nabla)_i \hat{A}_{\perp,m}(\bm{r},t), E_{\perp,m'} \left(\bm{r}^{\prime}\right)\left(\bm{r}^{\prime} \times \nabla^{\prime}\right)_j A_{\perp,m'}\left(\bm{r}^{\prime}\right)\right]\\
    =& i\hbar\varepsilon \int d^3r\int d^3r^{\prime} ( \hat{E}_{\perp,m} (\bm{r},t)(\bm{r} \times \nabla)_i \hat{A}_{\perp,m}(\bm{r},t) E_{\perp,m'} (\bm{r}')(\bm{r}' \times \nabla')_j A_{\perp,m'}(\bm{r}') \\
    &-E_{\perp,m'} (\bm{r}')(\bm{r}' \times \nabla')_j A_{\perp,m'}(\bm{r}') \hat{E}_{\perp,m} (\bm{r},t)(\bm{r} \times \nabla)_i \hat{A}_{\perp,m}(\bm{r},t) )\\
    =& =i \hbar \varepsilon_0 \int d^3r \int d^3r^{\prime}(\hat{E}_{\perp,m}(\bm{r} \times \nabla)_i \delta^\perp\left(\bm{r}-\bm{r}^{\prime}\right)\left(\bm{r}^{\prime} \times \nabla^{\prime}\right)_j A_{\perp,m'} \\
    & -E_{\perp,m'}(\bm{r}' \times \nabla')_j \delta^\perp\left(\bm{r}-\bm{r}^{\prime}\right)\left(\bm{r} \times \nabla\right)_i \hat{A}_{\perp,m} )\\
    =&i\hbar \int d^3r \hat{E}_{\perp,m}\varepsilon_{ijk}(r \times \nabla)_k \hat{A}_{\perp,m} = i\hbar \varepsilon_{ijk}\hat{L}_k \neq i\hbar \varepsilon_{ijk}(\hat{L}_k + \hat{S}_k),
\end{align*}
\begin{equation*}
    \left[\hat{L}_i, \hat{S}_{j}\right] = \varepsilon_0^2 \int d^3r\int d^3r^{\prime} [\hat{E}_{\perp,m} (\bm{r},t)(\bm{r} \times \nabla)_i \hat{A}_{\perp,m}(\bm{r},t), \epsilon_\textit{jpq} \hat{E}_{\perp,p} \hat{A}_{\perp,q}]=0 \neq i\hbar \varepsilon_{ijk}\hat{S}_k,
\end{equation*}
\noindent which is in stark contrast to \cite{Enk1994}. Thus, we arrive at the following commutation relation for the total AM operator:

\begin{equation}
    [\hat{J}_i,\hat{J}_j]=i\hbar \varepsilon_{ijk} \hat{L}_k \neq i\hbar\varepsilon_{ijk} \hat{J}_k
\end{equation}

\subsection{Rotation of an arbitrary field vector in SO(3)}

We address the question of whether the angular momentum operators generate rotations for a general vector field $V(\bm{r},t)$ in the SO(3) group. We decompose the longitudinal and transverse components of the field, given by:
\begin{equation}
    \bm{\hat{V}}^\parallel(\bm{r},t)= \int d^3k \mathcal{N}(\bm{k}) [\hat{a}_{\bm{k},3}\bm{\epsilon}(\bm{k},3)e^{i(\bm{k.r}-\omega t)} + h.c.],
\end{equation}
\begin{equation}
    \bm{\hat{V}}^\perp(\bm{r},t)= \int d^3k \mathcal{N}(\bm{k}) [\hat{a}_{\bm{k},-1}\bm{\epsilon}(\bm{k},-1)e^{i(\bm{k.r}-\omega t)} + \hat{a}_{\bm{k},+1}\bm{\epsilon}(\bm{k},1)e^{i(\bm{k.r}-\omega t)} + h.c.],
\end{equation}
where, $\mathcal{N}_{\parallel, \perp}(\bm{k})$ are the normalization constants. An infinitesimal rotation of the vector field by $\delta \theta$ along $j$ is given by:
\begin{align*}
    \hat{V}'_i=\hat{V}_i+\frac{i}{\hbar}[\hat{V}_i,\hat{\mathcal{R}}_j]\delta \theta_j
\end{align*}
where $\hat{\mathcal{R}}$ is a generator of rotations preserving the SO(3) group structure.Inspired from \cite{Mandel1982}, we realize that it is easier to compute the commutator $[\hat{J}_{i},\hat{V}_{j}^{\parallel}(\bm{r},t)]$, compared to individually verifying it for the OAM and the SAM. We first derive the commutators $[\hat{E}_{\perp,i},\hat{V}_{j}^{\parallel}(\bm{r},t)]$ and $[\hat{B}_{i},\hat{V}_{j}^{\parallel}(\bm{r},t)]$, and then use Equation \ref{eq1} to calculate the total commutator.


\begin{align*}
    &[\hat{E}_{\perp,i}(\bm{r},t),\hat{V}_{j}^{\parallel}(\bm{r}')]\\
    &=\left[i\int d^{3}k\sum_{\lambda=\pm1}\sqrt{\frac{\hbar\omega_{k}}{2\varepsilon_{0}}}\left(\hat{a}_{\bm{k},\lambda}\epsilon_{i}(\bm{k},\lambda)e^{i(\bm{k.r}-\omega t)}-h.c.\right),\int d^{3}k'\mathcal{N}(\bm{k}')\left(\hat{a}_{\bm{k}',3}\epsilon_{j}(\bm{k}',3)e^{i(\bm{k'.r'}-\omega't)}+h.c.\right)\right]\\
    &=i\sum_{\lambda=\pm1}\int d^{3}k\int d^{3}k'\mathcal{N}(\bm{k}')\sqrt{\frac{\hbar\omega_{k}}{2\varepsilon_{0}}}\left[\hat{a}_{\bm{k},\lambda},\hat{a}_{\bm{k}',3}^{\dagger}\right]\epsilon_{i}(\bm{k},\lambda)\epsilon_{j}^{*}(\bm{k}',3)e^{i(\bm{k.r}-\bm{k'.r'}-\omega t +\omega' t')}=0,
\end{align*}
since the commutator $\left[\hat{a}_{\bm{k},\lambda},\hat{a}_{\bm{k}',3}^{\dagger}\right]=0$. Similarly, we can show that $[\hat{B}_{i}(\bm{r},t),\hat{V}_{j}^{\parallel}(\bm{r}')]=0$ using the mode expansion of the magnetic field operator:
\begin{equation*}
    \hat{B}_{i}(\bm{r},t)=\frac{i}{c}\int d^{3}k\sqrt{\frac{\hbar}{2\omega_{k}\varepsilon_{0}}}\left(\left(\hat{a}_{\bm{k},1}\epsilon_i(\bm{k},2)-\hat{a}_{\bm{k},2}\bm{\epsilon}(\bm{k},1)\right)e^{i(\bm{k.r}-\omega t)}+h.c.\right).
\end{equation*}
Using Equation \ref{eq1}, we get:
\begin{align*}
    &[\hat{J}_i,\hat{V}^\parallel_j(\bm{r},t)]\\
    &=\varepsilon_0\int d^3r' r'_n\left[\hat{V}^\parallel_j(\bm{r},t), \hat{E}_{\perp,i}\hat{B}_n-\hat{E}_{\perp,n}\hat{B}_l\right]\\
    &=\varepsilon_0\int d^3r' r'_n\left(\hat{V}^\parallel_j(\bm{r},t)\hat{E}_{\perp,i}\hat{B}_n-\hat{V}^\parallel_j(\bm{r},t)\hat{E}_{\perp,n}\hat{B}_l - \hat{E}_{\perp,i}\hat{B}_n\hat{V}^\parallel_j(\bm{r},t)+\hat{E}_{\perp,n}\hat{B}_l \hat{V}^\parallel_j(\bm{r},t) \right)\\
    &=0,
\end{align*}
since $[\hat{E}_{\perp,i},\hat{V}_{j}^{\parallel}(\bm{r},t)]=[\hat{B}_{i},\hat{V}_{j}^{\parallel}(\bm{r},t)]=0$. Also, we can show from \cite{Mandel1982}:
\begin{equation*}
    [\hat{J}_i,\hat{V}_j^\perp(\bm(r))]=i\hbar\left( \varepsilon_{ijk}\hat{V}_k^\perp - \varepsilon_{ink}r'_n \nabla_k \hat{V}_j^\perp \right),
\end{equation*}

\section{Relativistic approach:}

We start with the Lagrangian density for the Maxwell field \cite{Yang2022,Yang2021}:
\begin{equation}
    \mathcal{L}_M = -\frac{1}{2\mu_0}(\partial_\mu A^\nu)(\partial^\mu A_\nu),
\end{equation}
where $\mu_0$ is the permeability in free space. This Lagrangian density is not a U(1) gauge-invariant quantity. The gauge-invariant standard Lagrangian $\mathcal{L}_{\textit{ST}} = -(1/4\mu_0)F^{\mu \nu}F_{\mu \nu}$ is related to the the Lagrangian for the Maxwell field by: 
\begin{equation}
    \mathcal{L}_{\textit{ST}}=\mathcal{L}_M + \frac{1}{2}(\partial^\mu A_\mu)^2 + \text{vanishing surface terms}.    
\end{equation}
The surface terms are dropped, as they vanish sufficiently fast at infinity. Thus, under Lorenz gauge condition $\partial_\mu A^\mu=0$, the two Lagrangian densities are equivalent. 

For our approach, we preserve the all the four degrees of freedom from $\mathcal{L}_M$, and then decompose the operators to remove the gauge-dependent components. We show that the gauge-dependent total AM operator is a generator of rotations under SO(3).

\subsection{Noether's theorem and Total AM:}
 An infinitesimally small transformation $\delta x$ is given by:
\begin{align*}
x'_{\mu} & =x_{\mu}+\delta x_{\mu}.
\end{align*}
Let us conside a vector field $A(x)$, which change as:
\[
A'(x')=A(x)+\delta A(x),
\]
and similarly, the Lagrange density transforms as:
\[
\mathcal{\mathcal{L}}'(x')=\mathcal{L}(x)+\delta L(x).
\]
We can also define a modified variation $\tilde{\delta}$ which addresses the change which is purely due to the field and not the coordinate shift.
\begin{align*}
\tilde{\delta}A(x) & =A'(x)-A(x)\\
 & =A'(x)-A'(x')+A'(x')-A(x)\\
 & =\delta A(x)-(A'(x')-A'(x))\\
 & =\delta A(x)-\frac{\partial A'(x)}{\partial x_{\mu}}\delta x_{\mu}\qquad\qquad\text{(first term of Taylor series expansion)}\\
 & =\delta A(x)-\frac{\delta A(x)}{\partial x_{\mu}}\partial x_{\mu}.
\end{align*}
From the invariance of action,
\[
\partial W=\int_{\Omega'}d^{4}x'\mathcal{L}'(x')-\int_{\Omega}d^{4}x\mathcal{L}(x)=0.
\]
Solving the above equation yields the equation of continuity:
\begin{align}
&\frac{\partial}{\partial x_{\mu}}f_{\mu}(x) =0,\\
\text{where,}\quad &f_{\mu}(x) =\frac{\partial\mathcal{L}(x)}{\partial(\partial^{\mu}A)}\delta A(x)-[\frac{\partial\mathcal{L}(x)}{\partial(\partial^{\mu}A)}\frac{\partial A}{\partial x^{\nu}}-g_{\mu\nu}\mathcal{L}(x)]\partial x^{\nu}
\end{align}

\noindent\textbf{Invariance under translation:} Here the shape of a field remains invariant under translation:
\[
A'^{\mu}(x')=A^{\mu}(x)+(\delta A^{\mu}(x)=0).
\]
From the Equation of Continuity,
\[
\frac{\partial}{\partial x_{\mu}}\big[\frac{\delta\mathcal{L}_{M}}{\partial(\partial^{\mu}A_{\sigma})}.\frac{\partial A_{\sigma}}{\partial x^{\nu}}-g^{\mu\nu}\mathcal{L}_{M}\big]=0
\]
From this equation, we obtain the canonical Energy-Momentum Tensor ($\Theta_{M}^{\mu\nu}$):
\begin{align}
\Theta_{M}^{\mu\nu} & =\frac{\delta\mathcal{L}_{M}}{\partial(\partial^{\mu}A_{\sigma})}.\frac{\partial A_{\sigma}}{\partial x^{\nu}}-g^{\mu\nu}\mathcal{L}_{M}\\ \nonumber
 & =-\frac{1}{\mu_{0}}(\partial^{\mu}A^{\sigma})(\partial^{\nu}A_{\sigma})+\frac{1}{2\mu_{0}}g^{\mu\nu}(\partial^{\rho}A^{\sigma})(\partial_{\rho}A_{\sigma})
\end{align}

\noindent \textbf{Invariance under rotation: }A general infinitesimal rotation can be written as:
\[
x'^{\mu}=x^{\mu}+(\delta\omega^{\mu\nu}x_{\nu}=\delta x^{\mu})
\]
The length of the vector can be obtained from the four-vector inner product:
\begin{align*}
x'^{\mu}x'_{\mu} & =(x^{\mu}+\delta\omega^{\mu\sigma}x_{\sigma})(x_{\mu}+\delta\omega_{\mu}^{\tau}x_{\tau})\\
 & =x^{\mu}x_{\mu}+\delta\omega^{\mu\sigma}x_{\sigma}x_{\mu}+\delta\omega_{\mu}^{\tau}x^{\mu}x_{\tau}\\
 & =x^{\mu}x_{\mu}+x_{\mu}x_{\nu}(\delta\omega^{\mu\nu}+\delta\omega^{\nu\mu})
\end{align*}
The length in Minkowski space is invariant under rotation $x'^{\mu}x'_{\mu}=x^{\mu}x_{\mu}$,
implying $\delta\omega^{\mu\nu}$ is anti-symmetric. Applying the rotational transformation on the vector field,
\[
A'^{\sigma}(x')=A^{\sigma}(x)+\delta A^{\sigma}(x)=A^{\sigma}(x)+\frac{1}{2}\delta\omega_{\mu\nu}(I^{\mu\nu})^{\sigma\rho}A^{\rho}(x).
\]
Here, $(I^{\mu\nu})^{\sigma\rho}=g^{\mu\sigma}g^{\nu\rho}-g^{\mu\beta}g^{\nu\sigma}$ is the infinitesimal generator of SO(3), which is anti-symmetric. There are 6 independent generators of SO(3): $(\mu,\nu)=(1,2),(1,3),(2,3)$ correspond to the 3 spatial rotations and $(\mu,\nu)=(0,1),(0,2),(0,3)$ corresponds to the 3 Lorentz boosts. From the Equation of Continuity,
\begin{align*}
f_{\mu}(x) & =\frac{\partial\mathcal{L}(x)}{\partial(\partial^{\mu}A_{\sigma})}\delta A_{\sigma}(x)-\Theta_{\mu\nu}\partial x^{\nu}\\
 & =\frac{\partial\mathcal{L}(x)}{\partial(\partial^{\mu}A_{\sigma})}\frac{1}{2}\delta\omega_{\mu\nu}(I^{\mu\nu})_{\sigma\rho}A_{\rho}(x)-\Theta_{\mu\nu}\delta\omega^{\nu\lambda}x_{\lambda},
\end{align*}
where,
\begin{align*}
\Theta_{\mu\nu}\delta\omega^{\nu\lambda}x_{\lambda} & =\frac{1}{2}.2\Theta_{\mu\nu}\delta\omega^{\nu\lambda}x_{\lambda}\\
 & =\frac{1}{2}(\Theta_{\mu\nu}\delta\omega^{\nu\lambda}x_{\lambda}-\Theta_{\mu\nu}\delta\omega^{\lambda\nu}x_{\nu})\\
 & =\frac{1}{2}\delta\omega^{\nu\lambda}(\Theta_{\mu\nu}x_{\lambda}-\Theta_{\mu\lambda}x_{\nu}).
\end{align*}
Substituting $\Theta_{\mu\nu}\delta\omega^{\nu\lambda}x_{\lambda}$ back in the expression of $f_{\mu}(x) $,
\[
f_{\mu}(x)=\frac{1}{2}\delta\omega^{\nu\lambda}M_{\mu\nu\lambda}(x),
\]
\[
M_{\mu\nu\lambda}=\Theta_{\mu\nu}x_{\lambda}-\Theta_{\mu\lambda}x_{\nu}+\frac{\partial\mathcal{L}(x)}{\partial(\partial^{\mu}A_{\sigma})}(I_{\nu\lambda})_{\sigma\tau}A_{\tau}.
\]
Here, $M_{\mu\nu\lambda}$ the angular momentum density tensor. For the Maxwell field,
\begin{align*}
    M_{M}^{\mu\nu\lambda} & =\Theta_{M}^{\mu\lambda}x^{\nu}-\Theta_{M}^{\mu\nu}x^{\lambda}+\frac{\partial\mathcal{L}_{M}(x)}{\partial(\partial_{\mu}A^{\sigma})}(I^{\nu\lambda})^{\sigma\tau}A_{\tau}\\
     & =\Theta_{M}^{\mu\lambda}x^{\nu}-\Theta_{M}^{\mu\nu}x^{\lambda}-\frac{1}{\mu_{0}}(\partial^{\mu}A_{\sigma})(g^{\nu\sigma}g^{\lambda\tau}-g^{\nu\tau}g^{\lambda\sigma})A_{\tau}\\
     & =\Theta_{M}^{\mu\lambda}x^{\nu}-\Theta_{M}^{\mu\nu}x^{\lambda}-\frac{1}{\mu_{0}}[(\partial^{\mu}A^{\nu})A^{\lambda}-(\partial^{\mu}A^{\lambda})A^{\nu}].
\end{align*}
To derive the angular momentum tensor, we perform a volume integral of the angular momentum density tensor $\mu=0$ ($\hat{\bm{\Tilde{J}}}\equiv M^{\nu\lambda}_M=\int d^4xM^{0,\nu\lambda}_M$):
\begin{align}
    \hat{\bm{\Tilde{J}}} & \equiv M_{M}^{\nu\lambda}=\int d^4x[\Theta_{M}^{0,\lambda}x^{\nu}-\Theta_{M}^{0,\nu}x^{\lambda}-\frac{1}{\mu_{0}}[(\partial^{0}A^{\nu})A^{\lambda}-(\partial^{0}A^{\lambda})A^{\nu}]]\\ \nonumber
     & =\frac{1}{c}\int d^3r[-\pi^{\mu}(\bm{r\times\nabla})A_{\mu}+\boldsymbol{\pi\times A}].
\end{align}
where, $r_i=x_{\nu=0,1}$ represent the spatial coordinates. One can verify that this tensor of angular momentum is a constant of motion and is an anti-symmetric quantity. The spin and the orbital components of the total AM are:
\begin{align}
    \hat{\bm{\Tilde{L}}}=&-\frac{1}{c}\int d^3r[\pi^{\mu}(\bm{r\times\nabla})A_{\mu}]\\
    \hat{\bm{\Tilde{S}}}=&\frac{1}{c}\int d^3r[\boldsymbol{\pi\times A}]
\end{align}

\subsection{Gauge decomposition and angular momentum commutation relations:}

We postulate the Equal Time Commutation Relations (ETCRs):
\begin{align*}
    [A^{\mu}(\bm{r},t),\pi^{\nu}(\bm{r}',t)] & =i\hbar cg^{\mu\nu}\delta^{3}\bm{r-r'}),\\
    [A^{\mu}(\bm{r},t),A^{\nu}(\bm{r}',t)] & =[\pi^{\mu}(\bm{r},t),\pi^{\nu}(\bm{r}',t)]=0,
\end{align*}
for the plane-wave expansion of the bosonic mode operators, given by:
\begin{align*}
    A^{\mu} & =\int d^{3}k\sum_{\lambda=0}^{3}\sqrt{\frac{\hbar}{2\varepsilon_{0}\omega\hat{a}_{\bm{k}}(2\pi)^{3}}}[\hat{a}_{\bm{k},\lambda}\epsilon^{\mu}(\boldsymbol{k},\lambda)e^{i\boldsymbol{k.x}}+h.c.],\\
    \pi^{\mu} & =i\int d^{3}k\sum_{\lambda=0}^{3}\sqrt{\frac{\hbar\omega\hat{a}_{\bm{k}}}{2\mu_{0}(2\pi)^{3}}}[\hat{a}_{\bm{k},\lambda}\epsilon^{\mu}(\boldsymbol{k},\lambda)e^{i\boldsymbol{k.x}}-h.c.].
\end{align*}
These creation and annihilation bosonic mode operators follow the commutation relations. To verify this, one can perform inverse transformation on the field operators to find the expression of the mode operators in terms of the field operators and compute the commutation relations.
\[
[\hat{a}_{\bm{k},\lambda},\hat{a}_{\bm{k}',\lambda'}^{\dagger}]=-g_{\lambda,\lambda'}\delta^{3}(\boldsymbol{k-k}'),\qquad[\hat{a}_{\bm{k},\lambda},\hat{a}_{\bm{k}',\lambda'}]=[\hat{a}_{\bm{k},\lambda}^{\dagger},\hat{a}_{\bm{k}',\lambda'}^{\dagger}]=0.
\]
We apply the ETCR commutation relations to derive the commutation relations for the AM operators:
\begin{align*}
    &[\hat{\Tilde{L}}_{M,i},\hat{\Tilde{L}}_{M,j}]\\
    & =(1/c)^{2}\int d^3r d^3r'[\pi^{p}(r\times\nabla)_{i}A^{p},\pi^{p'}(r'\times\nabla')_{j}A^{p'}]\\
     & =(1/c)^{2}\int d^3rd^3r'\big(\pi^{p}(r\times\nabla)_{i}A^{p}\pi^{p'}(r'\times\nabla')_{j}A^{p'}-\pi^{p'}(r'\times\nabla')_{j}A^{p'}\pi^{p}(r\times\nabla)_{i} A^{p}\big)\\
     & =(1/c)^{2}\int d^3rd^3r'\big(\pi^{p}(r\times\nabla)_{i}(i\hbar c g^{pp'}\delta^{3}(r-r'))(r'\times\nabla')_{j}A^{p'}-\pi^{p}(r\times\nabla)_{i}\pi^{p'}A^{p}(r'\times\nabla')_{j}A^{p'}\\
     & \quad-\pi^{p'}(r'\times\nabla')_{j}(i\hbar cg^{pp'}\delta^{3}(r-r'))(r\times\nabla)_{i}A^{p}+\pi^{p'}(r'\times\nabla')_{j}\pi^{p}A^{p'}(r\times\nabla)_{i}A^{p}\big)\\
     & =(1/c)^{2}\int d^3r\big(\pi^{p}(r\times\nabla)_{i}(i\hbar c)(r\times\nabla)_{j}A^{p}-\pi^{p}(r\times\nabla)_{j}(i\hbar c)(r\times\nabla)_{i}A^{p}\big)\\
     & =-\frac{i\hbar}{c}\int d^3r\big(\pi^{p}\varepsilon_{ijk}(r\times\nabla)_{k}A^{p}\big)=i\hbar\varepsilon_{ijk}\hat{\Tilde{L}}_{M,k},
\end{align*}

\begin{align*}
    &[\hat{\Tilde{S}}_{M,i},\hat{\Tilde{S}}_{M,j}]\\
    & =(1/c)^{2}\epsilon_{ikl}\epsilon_{jk'l'}\int d^3rd^3r'[\pi_{k}A_{l},\pi_{k'}A_{l''}]\\
     & =(1/c)^{2}\epsilon_{ikl}\epsilon_{jk'l'}\int d^3rd^3r'\big(\pi_{k}[A_{l},\pi_{k'}]A_{l''}-\pi_{k'}[\pi_{k},A_{l''}]A_{l}\big)\\
     & =(1/c)^{2}\epsilon_{ikl}\epsilon_{jk'l'}\int d^3rd^3r'\big(\pi_{k}i\hbar g^{k'l}\delta^{3}(r-r')A_{l''}-\pi_{k'}i\hbar g^{kl'}\delta^{3}(r-r')A_{l}\big)\\
     & =\frac{i\hbar}{c}\int d^3r[\epsilon_{ikl}\epsilon_{jll'}\pi_{k}A_{l'}-\epsilon_{ikl}\epsilon_{jkk'}\pi_{k}A_{l}]\\
     & =\frac{i\hbar}{c}\int d^3r[\epsilon_{ilk}\epsilon_{jkk'}\pi_{k'}A_{l'}-\epsilon_{ikl}\epsilon_{jll'}\pi_{k}A_{l'}]\\
     & =\frac{i\hbar}{c}\int d^3r[(\delta_{ij}\delta_{lk'}-\delta_{ik'}\delta_{lj})\pi_{k'}A_{l'}-(\delta_{ij}\delta \hat{a}_{\bm{k}l'}-\delta_{il'}\delta \hat{a}_{\bm{k}j})\pi_{k}A_{l'}]\\
     & =\frac{i\hbar}{c}\int d^3r[\varepsilon_{ijk}(\pi\times A)_{k}]=i\hbar\varepsilon_{ijk}\hat{\Tilde{S}}_{M,k}.
\end{align*}
Also, it is trivial to prove:
\begin{equation*}
    [\hat{\Tilde{L}}_i,\hat{\Tilde{S}}_j]=0.
\end{equation*}
\begin{equation}
    \therefore [\hat{\Tilde{J}}_i,\hat{\Tilde{J}}_j]=i\hbar\varepsilon_{ijk}\hat{\Tilde{J}}_k.
    \label{AM_gauge_dep}
\end{equation}
Equation \ref{AM_gauge_dep} suggests that we obtained the desired operator for the total AM as a generator of rotations in SO(3). But, one can check that under the $U(1)$ symmetric gauge transformation $A_\mu'(x) \rightarrow A_\mu(x) - \partial_\mu f(x)$, the total AM operator $\bm{\hat{J}}$ is not conserved. Moreover, we have four polarization degrees of freedom for the the electromagnetic field, meaning there exists two redundant degrees of freedom. From the plane-wave expansion of the Maxwell Hamiltonian,
\begin{equation*}
    \mathcal{H}_M=\int d^3k \hbar \omega_k (\hat{a}_{\bm{k},1}^\dagger \hat{a}_{\bm{k},1} + \hat{a}_{\bm{k},2}^\dagger \hat{a}_{\bm{k},2} + \hat{a}_{\bm{k},3}^\dagger \hat{a}_{\bm{k},3} - \hat{a}_{\bm{k},0}^\dagger \hat{a}_{\bm{k},0}).
\end{equation*}
The Maxwell Hamiltonian contains the scalar photon $\lambda=0$ with a negative frequency. We use Helmholtz decomposition to separate the rotational (transverse) and non-rotational (longitudinal) components of the field operators and apply a similar approach as Section A to remove the pure gauge-dependent terms from the AM operators.
\begin{equation}
    \hat{\bm{\Tilde{L}}}=-\frac{1}{c}\int d^3r[\pi_\perp^{j}(\bm{r\times\nabla})A_{\perp}^j+\pi_\parallel^{j}(\bm{r\times\nabla})A_{\parallel}^j-\pi_0(\bm{r\times\nabla})A_0],
\end{equation}
\begin{equation}
    \hat{\bm{\Tilde{S}}}=\frac{1}{c}\int d^3r [\boldsymbol{\pi_\perp\times A_\perp}+\boldsymbol{\pi_\parallel\times A_\perp}+\boldsymbol{\pi_\perp\times A_\parallel}+\boldsymbol{\pi_\parallel \times A_\parallel}].
\end{equation}

Since $A_\parallel$ is purely gauge-dependent and $A_0$ corresponds to the scalar photon, the gauge-independent quantum AM operators are \cite{Yang2022}:
\begin{equation}
    \bm{\hat{L}} =-\frac{1}{c}\int d^3r[\pi_\perp^{j}(\bm{r\times\nabla})A_{\perp}^j]=-i\hbar \int d^3k \sum_{\lambda=1,2}[\hat{a}_{\bm{k},\lambda}^\dagger(\boldsymbol{k \times \nabla_k})\hat{a}_{\bm{k},\lambda}],
    \label{OAM_obs}
\end{equation}
\begin{equation}
    \bm{\hat{S}} =\frac{1}{c}\int d^3r [\boldsymbol{\pi_\perp\times A_\perp}]=i\hbar \int d^3k[\hat{a}_{\bm{k},2}^\dagger \hat{a}_{\bm{k},1} - \hat{a}_{\bm{k},1}^\dagger \hat{a}_{\bm{k},2}]\boldsymbol{\epsilon}(\boldsymbol{k},3).
    \label{SAM_obs}
\end{equation}
From Equations \ref{OAM_obs} and \ref{SAM_obs}:
\begin{align*}
    [\hat{\Tilde{L}}_{i},\hat{\Tilde{L}}_{j}] & =(1/c)^{2}\int d^3rd^3r'[\pi^{p}(r\times\nabla)_{i}A^{p},\pi^{p'}(r'\times\nabla')_{j}A^{p'}]\\
     & =-\hbar^{2}\int d^{3}kd^{3}k'\sum_{\lambda,\lambda'=1}^{3}[\hat{a}_{\bm{k},\lambda}^{\dagger}(k\times\nabla_k)_{i}\hat{a}_{\bm{k},\lambda},\hat{a}_{\bm{k}',\lambda'}^{\dagger}(k'\times\nabla_{k'} )_{j}\hat{a}_{\bm{k}',\lambda'}]=i\hbar \hat{\Tilde{L}}_{k}\\
    \Rightarrow[\hat{L}_{i},\hat{L}_{j}] & =-\hbar^{2}\int d^{3}kd^{3}k'\sum_{\lambda,\lambda'=1,2}[\hat{a}_{\bm{k},\lambda}^{\dagger}(k\times\nabla_k)_{i}\hat{a}_{\bm{k},\lambda},\hat{a}_{\bm{k}',\lambda'}^{\dagger}(k'\times\nabla_{k'})_{j}\hat{a}_{\bm{k}',\lambda'}]\\
    &=i\hbar \hat{L}_{k},\quad\text{(follows similar calculations).}
\end{align*}
The derivations for the commutation relations of SAM are trivial. Hence, we obtain:
\begin{align}
    [\hat{L}_i ,\hat{L}_j ]=&i\hbar \varepsilon_{ijk}\hat{L}_k, \\
    [\hat{S}_i ,\hat{S}_j ]=&0,\\
    [\hat{L}_i ,\hat{S}_j ]=&0.
\end{align}
Using the commutation relations derived above, we finally obtain the correct commutation relation for the gauge-independent total AM operator as:
\begin{equation}
    [\hat{J}_i ,\hat{J}_j ]=i\hbar \varepsilon_{ijk}\hat{L}_k  \neq i\hbar \varepsilon_{ijk}\hat{J}_k. 
\end{equation}

\section{Gauge-dependent quantum AM operators as the generator of rotations}
We separately derive the commutators of the OAM and the SAM with the gauge field, and sum their cumulative products to obtain the total AM commutator with the field vector. For our calculations, we use the reduced form of the gauge-dependent OAM and SAM operators derived from Noether's theorem in \cite{Yang2021}.
\begin{align*}
    &[\hat{\tilde{L}}_{m},\hat{V}_{n}]\\
    &=-\frac{i\hbar}{2}\sum_{\lambda,\lambda'}\int d^{3}k\int d^{3}k'\mathcal{N}(k')\left[\left(\hat{a}_{\bm{k},\lambda}^{\dagger}(\bm{k}\times\boldsymbol{\nabla}_{k})_{m}\hat{a}_{\bm{k},\lambda}-\hat{a}_{\bm{k},\lambda}(\bm{k}\times\boldsymbol{\nabla}_{k})_{m}\hat{a}_{\bm{k},\lambda}^{\dagger}\right),(\hat{a}_{\bm{k'},\lambda'}\epsilon_{n}(\bm{k'},\lambda')e^{i(\bm{k'.r}-\omega' t)}+h.c.)\right]\\&=-\frac{i\hbar}{2}\sum_{\lambda,\lambda'}\int d^{3}k\int d^{3}k'\mathcal{N}(k')\left[\hat{a}_{\bm{k},\lambda}^{\dagger}(\bm{k}\times\boldsymbol{\nabla}_{k})_{m}\hat{a}_{\bm{k},\lambda},(\hat{a}_{\bm{k'},\lambda'}\epsilon_{n}(\bm{k'},\lambda')e^{i(\bm{k'.r}-\omega' t)}+h.c.)\right]\\&\qquad+\frac{i\hbar}{2}\sum_{\lambda,\lambda'}\int d^{3}k\int d^{3}k'\mathcal{N}(k')\left[\hat{a}_{\bm{k},\lambda}(\bm{k}\times\boldsymbol{\nabla}_{k})_{m}\hat{a}_{\bm{k},\lambda}^{\dagger},(\hat{a}_{\bm{k'},\lambda'}\epsilon_{n}(\bm{k'},\lambda')e^{i(\bm{k'.r}-\omega' t)}+h.c.)\right]\\&=-\frac{i\hbar}{2}\sum_{\lambda,\lambda'}\int d^{3}k\int d^{3}k'\mathcal{N}(k')\left[\hat{a}_{\bm{k},\lambda}^{\dagger}\epsilon_{mrs}k_{r}\frac{\partial }{\partial k_{s}}\hat{a}_{\bm{k},\lambda},(\hat{a}_{\bm{k'},\lambda'}\epsilon_{n}(\bm{k'},\lambda')e^{i(\bm{k'.r}-\omega' t)}+h.c.)\right]\\&\qquad+\frac{i\hbar}{2}\sum_{\lambda,\lambda'}\int d^{3}k\int d^{3}k'\mathcal{N}(k')\left[\hat{a}_{\bm{k},\lambda}\epsilon_{mrs}k_{r}\frac{\partial }{\partial k_{s}}\hat{a}_{\bm{k},\lambda}^{\dagger},(\hat{a}_{\bm{k'},\lambda'}\epsilon_{n}(\bm{k'},\lambda')e^{i(\bm{k'.r}-\omega' t)}+h.c.)\right]\\&=-\frac{i\hbar}{2}\sum_{\lambda,\lambda'}\int d^{3}k\int d^{3}k'\mathcal{N}(k')\hat{a}_{\bm{k},\lambda}^{\dagger}\epsilon_{mrs}k_{r}\frac{\partial }{\partial k_{s}}\hat{a}_{\bm{k},\lambda}(\hat{a}_{\bm{k'},\lambda'}\epsilon_{n}(\bm{k'},\lambda')e^{i(\bm{k'.r}-\omega' t)}+h.c.)\\&\qquad+\frac{i\hbar}{2}\sum_{\lambda,\lambda'}\int d^{3}k\int d^{3}k'\mathcal{N}(k')\hat{a}_{\bm{k},\lambda}\epsilon_{mrs}k_{r}\frac{\partial }{\partial k_{s}}\hat{a}_{\bm{k},\lambda}^{\dagger}(\hat{a}_{\bm{k'},\lambda'}\epsilon_{n}(\bm{k'},\lambda')e^{i(\bm{k'.r}-\omega' t)}+h.c.)\\
    &=i\hbar\sum_{\lambda,\lambda'}\int d^{3}k\int d^{3}k'\mathcal{N}(k')\left[\hat{a}_{\bm{k},\lambda},\hat{a}_{\bm{k},\lambda}^{\dagger}\right]\epsilon_{mrs}k_{r}(\hat{a}_{\bm{k'},\lambda'}\frac{\partial }{\partial k_{s}}\left(\epsilon_{n}(\bm{k'},\lambda')e^{i(\bm{k'.r}-\omega' t)}\right)+h.c.)\\
    &=i\hbar\sum_{\lambda}\int d^{3}k\mathcal{N}(k)\left(\hat{a}_{\bm{k},\lambda}\epsilon_{mrs}k_{r}\frac{\partial }{\partial k_{s}}\left(\epsilon_{n}(\bm{k},\lambda)e^{i(\bm{k.r}-\omega t)}\right)+h.c.\right)\\
    &=i\hbar\epsilon_{mrs}r_{s}\frac{\partial }{\partial r_{r}}\hat{V_{n}}.
\end{align*}

The quantized form of the gauge-dependent SAM operator is given by:
\begin{equation*}
    \hat{\tilde{\bm{S}}}=i\hbar\int d^{3}k\sum_{\lambda=1}^{3}\hat{\bm{s}}_{k,\lambda},
\end{equation*}
where
\begin{align*}
    &\hat{\bm{s}}_{k,1}=(\hat{a}_{\bm{k},3}^{\dagger}a_{\bm{k},2}-\hat{a}_{\bm{k},2}^{\dagger}\hat{a}_{\bm{k},3})\bm{\epsilon}(\bm{k},1)\\
    &\hat{\bm{s}}_{k,2}=(\hat{a}_{\bm{k},1}^{\dagger}\hat{a}_{\bm{k},3}-\hat{a}_{\bm{k},3}^{\dagger}\hat{a}_{\bm{k},1})\bm{\epsilon}(\bm{k},2)
    \\&\hat{\bm{s}}_{k,3}=(\hat{a}_{\bm{k},2}^{\dagger}\hat{a}_{\bm{k},1}-\hat{a}_{\bm{k},1}^{\dagger}\hat{a}_{\bm{k},2})\bm{\epsilon}(\bm{k},3)
\end{align*}
It is trivial to prove:
\begin{align*}
    [\hat{\tilde{S}}_{m},\hat{V}_{n}]=i\hbar\epsilon_{mnr}\hat{V_{r}}
\end{align*}
Adding the OAM and the SAM commutation relations, we get:
\begin{align*}
    [\hat{\tilde{J}}_{m},\hat{V}_{n}]=i\hbar\left( \epsilon_{mnr}\hat{V_{r}}-\epsilon_{mrs}r_{s}\frac{\partial }{\partial r_{r}}\hat{V_{n}} \right)
\end{align*}

We perform an infinitesimally small unitary rotation of the gauge field $\hat{V}(\bm{r},t)$ by $\delta \bm{\theta}$ and get the transformed vector $\hat{V}'_{i}(\bm{r},t)$:
\begin{align*}
    &\hat{V}'(\bm{r},t)=e^{i\hat{\tilde{\bm{J}}} \cdot \delta\bm{\theta}/\hbar}\hat{V}(\bm{r},t)e^{-i\hat{\tilde{\bm{J}}} \cdot \delta\bm{\theta}/\hbar}\\
    \Rightarrow& \hat{V}'_i(\bm{r},t)=\hat{V}(\bm{r},t)+\frac{i}{\hbar}[\hat{\tilde{J}}_{j},\hat{V}_{i}(r)]\delta\theta_{j} + \mathcal{O}(\delta \theta^2).
\end{align*}
Since $\delta \bm{\theta}$ is infinitesimally small, we assume that the higher order terms vanishes. We have:
\begin{align*}
    \hat{V}'_{i}(\bm{r},t)&=\hat{V}_{i}(\bm{r},t)+\frac{i}{\hbar}[\hat{\tilde{J}}_{j},\hat{V}_{i}(r)]\delta\theta_{j}=\hat{V}_{i}+\epsilon_{ijk}\delta\theta_{j}\hat{V}_{k}-\epsilon_{imn}\delta\theta_{j}r_{n}\nabla_{m}\hat{V}_{i}\\
    &=\hat{V}(\bm{r},t)+\delta\bm{\theta}\times\hat{\bm{V}}-\left[\delta\bm{\theta}\times\bm{r}\right]\cdot\boldsymbol{\nabla}\hat{\bm{V}}(\bm{r},t)\\
    &=\delta\bm{\theta}\times\hat{\bm{V}}+\left(1-\left[\delta\bm{\theta}\times\bm{r}\right]\cdot\boldsymbol{\nabla}\right)\hat{\bm{V}}(\bm{r},t).
\end{align*}
To the first-order of $\delta \bm{\theta}$,
\begin{align*}
    &\left(1-\left[\delta\bm{\theta}\times\bm{r}\right]\cdot\boldsymbol{\nabla}\right)\hat{\bm{V}}(\bm{r},t)=\hat{V}(\bm{r}-\delta{\bm{\theta}}\times\bm{r},t).
\end{align*}
Also, under this first-order expansion, using the Taylor series expansion we have: $\delta\bm{\theta}\times\hat{V}(\bm{r}-\delta{\bm{\theta}}\times\bm{r},t)=\delta\bm{\theta}\times\hat{\bm{V}}(\bm{r},t)$. Therefore, we obtain relation for the rotated field vector $\hat{V}'$, given by:
\begin{align*}
    \hat{V}'(\bm{r},t)=\left(1+\delta\bm{\theta}\times\right)\hat{\bm{V}}(\bm{r}-\delta{\bm{\theta}}\times\bm{r},t),
\end{align*}
thus explicitly proving that the total AM operator is a generator of rotations \cite{Mehta1968}.




\bibliography{biblio}




\end{document}